# A geothermal hydro wind PV hybrid system with energy storage in an extinct volcano for 100% renewable supply in Ometepe, Nicaragua

Fausto A. Canales[1], Jakub K. Jurasz[2,3] and Alexandre Beluco[4,*]

[1] Universidad de la Costa, Department of Civil and Environmental, Barranquilla, Atlántico, Colombia; fausto.canales.v@gmail.com
[2] AGH University, Faculty of Management, Department of Engineering Management, Kraków, Poland
[3] MDH University, School of Business Society and Engineering, Västerås, Sweden; jakubkamiljurasz@gmail.com
[4] Universidade Federal do Rio Grande do Sul, Instituto de Pesquisas Hidráulicas, Porto Alegre, Rio Grande do Sul, Brazil; albeluco@iph.ufrgs.br
* Correspondence: fausto.canales.v@gmail.com (F.A.C.); albeluco@iph.ufrgs.br (A.B.)

**Abstract:** Renewable resources are constantly increasing their share in energy systems around the world. This paper evaluates how the capital cost of renewable technologies affects the optimal configuration and cost of energy of an isolated power system, comprising only renewable resources. HOMER software was adapted to include and simulate pumped storage hydropower and geothermal power plants. Ometepe island, Nicaragua, was selected as case study because wind, solar and geothermal resources are available, but more importantly, it has an extinct volcano with a crater lake on its top that could be used as the upper reservoir for pumped storage hydropower. When geothermal is considered, the results show that this technology is able to serve the base load of the system, reducing the required installed capacity of other resources, as well as decreasing the storage requirements and excess electricity production. When the geothermal option is not included, the low complementarity in time of the other variable resources increases the required size of the solar and wind parks , amounting to up to 6.5 times the peak power, consequently rising the cost of energy and excess electricity production. The different system configuration results demonstrated that economic aspects of renewable generation are at least as important as the natural resources availability.

## Abbreviations, Acronyms, and Symbols used in this paper

- AC: Alternating Current
- $C_B$: Battery capacity, in Ampere·hour [Ah]
- CO2: Carbon dioxide
- COE: Cost of electricity, in $/kWh
- $E_S$: Stored energy, in kWh
- GWh/y: Gigawatt hour per year
- $gCO_2/kWh$: Grams of carbon dioxide per kilowatt hour
- H: Effective head, in meters [m]
- HOMER: Hybrid Optimization Model for Electric Renewables
- I: Electric current, in Amperes [A]
- kV: Kilovolts
- kWh: Kilowatt hour
- m.a.s.l.: Meters above sea level
- MERRA-2: Modern-Era Retrospective Analysis for Research and Applications, Version 2
- MW: Megawatt
- MWh: Megawatt hour
- NASA: National Aeronautics and Space Administration
- NPC: Net present cost

- O&M: Operation and Maintenance
- $P_{bat}$: Electric power delivered by the battery, in kilowatts [kW]
- PSH: Pumped Storage Hydropower
- PV: Photovoltaic
- $Q_P$: Pump flow
- RES: Renewable Energy Sources
- SIN: National Interconnected System
- VRES: Variable Renewable Energy Sources
- V: Voltage, in volts [V]
- Vol: Active volume of a reservoir, in cubic meters [$m^3$]
- y: Year
- Z: Altitude, in m
- $\eta_{hyd}$: Conversion efficiency, %

**Content**

**1. Introduction**

Non-combustion renewable energy sources (solar, wind, hydropower, and geothermal) are becoming the pillars of electric power systems around the world. The goal of achieving an electric power system based entirely on renewables, besides been investigated by numerous authors [1 -6], has also become a reality over the recent years in many areas across the globe. The following locations (cities, regions, islands) are examples where renewables penetration reached 100%: Aspen, Colorado, in the United States, which in 2015 removed coal plants and now it covers its energy demand with hydroelectricity, wind power, photovoltaics an geothermal energy [7]; Tokelau islands, administered by New Zealand, use photovoltaic systems with battery backup [8]; Costa Rica, which in 2016 and 2017 had its electricity produced entirely using renewables for over 300 days, and set a goal of becoming carbon-neutral within this century [9]; the Danish island of Samso, with a negative carbon footprint, uses biomass and wind power to serve its load, and is connected to the mainland for energy export and balancing [10]. The list is obviously not complete, but the growing number of communities supplied entirely from renewables tread the new ground and prove that such situation is possible.

The aforementioned examples confront a common notion, described in [11], that variable renewable energy sources (VRES), like wind and solar power, are unreliable and intermittent to a degree that they are unable to contribute significantly to electric power supply or to serve baseload power. To overcome the intermittency and stochastic nature of these energy sources, autonomous power systems with high levels of these renewables need adequate actions to balance energy supply and demand, in order to main-tain the electricity supply quality by controlling grid frequency and voltage. To achieve this balance, one of the most common approaches is the combination of dispatchable technologies and energy storage.

Geothermal power and hydropower reservoirs are the two non-combustion renewables capable of providing dispatchable generation. With a capacity factor between 45% and 90%, geothermal energy is not heavily affected by weather like solar, wind or even hydro; and it can serve both base load and peak loads. However, it is generally more economical to use geothermal plants to supply base load [12].

Pumped storage hydropower (PSH) is a mature and efficient form of bulk energy storage which has drawn significant attention as a viable option to facilitate the integration of VRES in isolated areas and national power grids [13, 14]. One the most important costs associated to hydropower with storage is the one related to building the reservoirs. Specifically, for PSH, the cost per kilowatt·hour (kWh) for a PSH plant that already has the two reservoirs available is around 18% of the cost of a PSH plant that requires two new reservoirs, and 40% of a system including one new reservoir [15]. In general, PSH can present various configurations: both reservoirs are man-made (e.g.: Goldisthal Pumped-Storage Plant, Germany); natural lakes connected by canal (e.g.: Żydowo Pumped Storage Hydroelectric Power Plant, Poland); one of the reservoirs is a natural water body and the second one is an artificial reservoir (e.g.: Żarnowiec Pumped Storage Power Station, Poland); the lower reservoir is using salt water from the ocean (e.g.: Okinawa Yanbaru Seawater Pumped Storage Power Station, Japan); PSH using exploited (abandoned) deep mines [16, 17]. Despite many possible configurations of PSH plants, the main factor determining suitability of given site for PSH development is a beneficial ratio of vertical (head) to horizontal distance between reservoirs [18].

The sustainability of communities living on islands and in remote communities across the world has been directly tied to fuel oil availability and their increasing price. This commodity propels ships and trucks that allow transporting goods into and from the island; fuels the vehicles that allow tourism and mobilization within these areas, and, perhaps most important, has enabled these communities to consistently generate electricity [19].

There is a rich body of literature on hybrid systems based on VRES coupled with PSH operating on islands. We briefly introduce some of those works written over the last years. Duić et al. [20] showed how PSH can be used to increase VRES penetration in Porto Santo island. Bueno and Carta [21] presented a techno-economic analysis of wind powered PSH for El Hierro island. Their follow-up study investigated how wind-PSH hybrid power systems could increase the penetration of renewables in the Canary Islands [22]. Caralis and Zervois [23] analyzed the operation of a wind-PSH project applied to Greek islands, and in their next study [24] they assessed the market potential of wind-PSH for autonomous power systems on Islands. Papaefthimiou et al. [25] presented operating policies for wind-PSH stations for island grids. Katsaprakakis et al., [26] proposed a wind-PSH system for an insular power system for the islands Karpathos and Kasos. Kapsali et al., [27] performed a sensitivity analysis of wind-PSH from the economic perspective, using the Greek islands as case study. Ma et al. [28] considered adding photovoltaic (PV) modules to a wind-PSH hybrid power system and analyzed it from a technical perspective, for a remote island in Hong Kong. Papaefthymiou and Papathanassiou [29] presented an optimal sizing method for wind-PSH stations for insular systems. A similar approach was presented by Ma et al., [30] aiming the optimal design of an autonomous solar-wind-PSH station. Barreira et al. [31] considered an off-stream PSH project as an option to increase VRES penetration in Santiago island (Galápagos). Tsamaslis et al. [32] showed a concept of hybridizing PVs with PSH to increase VRES penetration and achieve grid benefits for Cyprus.

The most extensively used software tool in research studies associated to hybrid power systems is HOMER (Hybrid Optimization Model for Electric Renewables) [33]. HOMER can be described as a computer model that assists in the prefeasibility and design process of hybrid power systems involving different energy sources (PV modules, wind turbines, run-of-river hydropower, batteries, generators, etc.). HOMER is primarily an economical model which is able of modeling the physical behavior and lifecycle cost of a power system, in order to select the best system configuration. HOMER is able to perform three main tasks: simulation, optimization and sensitivity analysis.

A limitation of HOMER is that it does not include the features to directly model some renewable technologies, among these, PSH and Geothermal. Nevertheless, using some equivalent representations and considerations, as explained in Canales and Beluco [34] and the user support available in the HOMER energy website [35], the user is able to use HOMER for assessing a hybrid power system that considers PSH and geothermal options. Upcoming HOMER releases or custom versions could include specific elements to model these technologies, but the Legacy Version, universally available and used in this study, does not have these tools.

Ometepe island, in Nicaragua, Central America, was selected as case study for this paper because it presents some interesting features for evaluating the method and assessing the technical feasibility of a 100% non-combustion RES system: a) it has a crater lake at the top of an extinct volcano that could be used as the upper reservoir of a PSH plant; b) it is located in a large tropical fresh water lake that could act as the lower reservoir of the PSH system; c) the island has identified geothermal potential, d) solar and wind energy resources are available.

The recent paper by Meza et al.[36] also works with the idea of reaching 100% of the energy supply to the island of Ometepe with renewable resources. Relatively detailed survey of the wind and solar potentials are presented and the possibility of energy storage in the crater of the Maderas volcano is also considered. Geothermal energy is mentioned, but according to the authors investments in research and development are still required. This work is focused on surveying the available renewable resources, also discussing their possible complementarity, and comparing them with consumption of the island's population, suggesting actions to reduce or optimize energy consumption.

Within this context, the aim of this paper is to evaluate how the capital cost of renewable technologies affects the optimal configuration and cost of electricity of an isolated electric power system, comprising only non-combustion renewables (geothermal, wind, solar, PSH). Some of the merits of this paper related to filling the gap in the literature are: 1) It shows how to simultaneously include geothermal power and PSH in a HOMER model, along with other VRES; 2) The method presented here, and similar data sources can be adapted to other places around the world with similar geographic characteristics, especially regarding the PSH potential of extinct crater lakes. The structure of this paper, following this brief introduction, includes a description of the method adopted and the presentation of the case study (a hybrid system on the island of Ometepe, Nicaragua); after that, the results are presented and discussed, closing with the main conclusions of this work.

## 2. Method

This paper evaluates how the capital cost of wind turbines and PV panels affects the system configuration of a 100% non-combustion renewable isolated system, using PSH as energy storage system, and considering the impact of geothermal power. The Legacy Version of HOMER Energy software [37] is used as the main working tool, because it allows the user to simulate, optimize and conduct sensitivity analysis on hybrid systems including different renewable energy resources and PSH, using some useful modifications explained in this section.

### 2.1. Pumped Storage Hydropower modeling in HOMER

Just like batteries, hydropower reservoirs can store energy and supply it according to the load demands. Based on this similarity, Canales and Beluco [34] proposed a method for allowing representing PSH as an equivalent battery in HOMER (with properties that remain constant during its lifetime), using a converter to represent the installed capacity of the hydropower plant. The following equation can be used to estimate the total stored energy $E_S$ [kWh] in the active volume of a reservoir Vol [m³]:

$$E_S = \frac{9.81 \times \eta_{hyd} \times H \times Vol}{3600} \tag{1}$$

In Eq. (1), H is the effective head [m], which is considered constant, and $\eta_{hyd}$ is the conversion efficiency of the electromechanical equipment [%].

Similarly, and assuming a fixed voltage V [volts] and capacity $C_B$ [Ampere·hour, Ah], the following equation can be used to calculate the stored energy in the equivalent battery representing the hydropower reservoir:

$$E_S = \frac{V \times C_B}{1000} \tag{2}$$

The electric power delivered by the battery $P_{bat}$, measured in kilowatts [kW], is proportional to the product of electric current I in Amperes [A] and the voltage V, and can be calculated by the expression:

$$P_{bat} = \frac{I \times V}{1000} \tag{3}$$

HOMER finds the optimal configuration in terms of the total net present cost (NPC) of the system, including the size of the battery bank, which can be translated as the optimal size of the active volume of the upper reservoir. The unitary equivalent battery represents the minimum active volume of the upper reservoir, therefore, for the method presented in Canales and Beluco [34], the quantity of batteries are direct multiples of the reference volume.

As explained in [38], the method assumes asynchronous electromechanical equipment (pumps and turbines), enabling the equipment to adjust the rotation speed and regulate the amount of energy absorbed or delivered, allowing to reduce the number of starts and stops of the machinery. However, it is common practice to use rectifiers and multiple units of synchronous equipment. More details and considerations to use this method are described in [34] and [38].

### 2.2. Geothermal modeling in HOMER

According to the directions described in [35] and also based on the considerations in [34], geothermal modeling in HOMER requires creating a new AC (Alternating Current) generator. It is also necessary to define a new fuel representing the properties of the geothermal resource, being the most important, the fuel carbon content. The carbon dioxide ($CO_2$) emissions from geothermal fields in tropical areas range from 13 to 380 g/kWh, much lower than for natural gas (453 g/kWh), oil (906 g/kWh) and coal (1042 g/kWh) [12]. Because of the nature of the geothermal resource, the fuel cost is set to zero in HOMER, but the Operation and Maintenance (O&M) costs are expressed in US$/hr. In this paper we want to assess if the capital cost and lack of complementarity between the different available VRES affect the optimal configuration of the system. Because of this, the minimum load ratio in HOMER is set to 100%, indicating that the generator is considered to run only at full blast and serve base load. For the entire period, the operating mode is set to "optimized" in HOMER.

### 2.3. Sensitivity analysis, optimal system configuration and simulation with HOMER

One of the main features of HOMER is the capacity to perform sensitivity analysis under a range of uncertainties or/and assumptions in the model inputs. This capacity, along with its user-friendly interface, detailed documentation and solidity proven by its thousands of users around the world [33], is the reason why the authors of this paper selected HOMER as the working tool to perform the analysis and simulation of the system described in the case study.

The preliminary assessment and design of hybrid power systems usually involves a large number of options and uncertainties related to key parameters like the load size, stochastic behavior of renewable sources, storage or backup power requirements, etc., thus making the process a difficult task. Particularly for systems including intermittent renewables sources, like wind and solar radiation, HOMER is able to overcome these challenges by running simulation in one-hour time steps, allowing acceptable accuracy by capturing the most important features of how the load demands are served by these VRES, without demanding excessive computation resources that would make the optimization and sensitivity analysis impractical [39].

HOMER is able to simulate, optimize and conduct sensitivity analysis for many system configurations at one run, and for each one of them and based on the set of constraints, the model evaluates the performance of the system for each hour of the year, determining its technical feasibility and estimating its life-cycle cost. The hybrid system proposed in this study is represented in the diagram of Figure 1.

In the optimization process, the NPC is the measure used by HOMER to rank the feasible system configurations, prioritizing from the most economic to the most expensive. The NPC includes initial capital cost, replacements, O&M, fuel and the cost of purchasing electricity from the grid within the project lifetime, discounted to the present. More detailed information about HOMER is available at [37] and [39].

The sensitivity analysis of this paper tries to offer insight of how HOMER can assess the impact of renewable energy capital costs on the systems configuration in a scenario of 100% non-combustion renewables, with solar and wind as the VRES. It will also evaluate the effect of considering geothermal power to serve baseload.

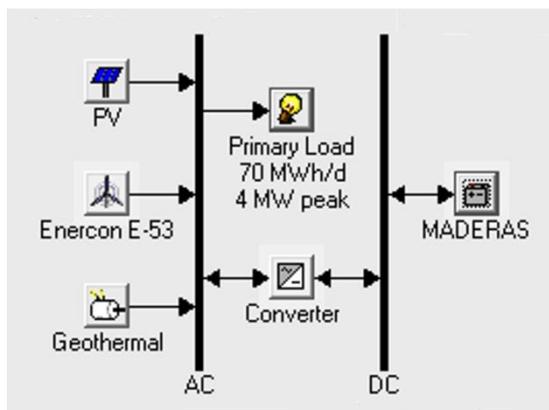

**Figure 1.** Schematic diagram of the proposed hybrid system for Ometepe Island. 'MADERAS', 'Converter' and DC bus bar jointly simulate the use of the Maderas' crater lake as an upper reservoir in a PSH plant.

Additionally, HOMER is able to estimate the harmful pollutant emissions that could be reduced by shifting to power systems with high percentages of non-combustion renewables. Besides compelling developed countries to apply environmental friendly energy policies, the common global agenda is to help developing countries to reduce their $CO_2$ and other greenhouse gases emissions, without thwarting their economic progress [40]. This paper uses information available at the electricity map site [41] related to $CO_2$ estimates of electricity consumption for several countries around the world, and the following approach is adopted: 1) The average $CO_2$ emissions (including all energy sources) produced by the national power grid during a 24 hour period were calculated and used in HOMER to estimate the $CO_2$ emissions that the demand would produce if all electricity was provided by the national grid; 2) when geothermal is being considered, the reference values of the $CO_2$ emissions are used to calculate the emissions of the 100% non-combustion renewable power system , allowing to compare and estimate the $CO_2$ emissions that might be reduced by implementing this hybrid power system.

**3. A hybrid system in the island of Ometepe**

This section describes the renewable resources, loads and components considered in this case study and included in the HOMER simulation to evaluate an autonomous hybrid power system to serve the electricity demands at Ometepe island, Nicaragua. By 2014, non-combustion renewables accounted for ~35.2% of the installed capacity of the National Interconnected System (Sistema Interconectado Nacional – SIN) of this Central American country [42].

In order to keep the quality of the electricity supply, power systems based on high shares of renewables require backup power to balance the intermittent and fluctuating behavior of sources like wind and sun radiation. In Nicaragua, this balance is provided by fuel generators. However, in this paper, a 100% renewable system is assessed, by using PSH instead of fuel to supply electricity in periods of high demand. Table 1 presents the search space used in HOMER for the case study.

In order to reduce infrastructure investment, PSH frequently uses exhausted mines or natural water reservoirs. For this case study, the crater lake at the top of Maderas, an extinct volcano, and Lake Nicaragua, a large tropical fresh water lake, are used as upper and lower reservoirs of the PSH plant. The island of Ometepe is formed by two volcanoes and is located in Lake Nicaragua, in the southwestern part of Nicaragua [43].

Ometepe has an area of 276 km², about 31 km long by 5 km to 10 km wide, housing a population of just over 40,000 inhabitants. Ometepe is formed by two volcanoes:Concepcion, reaching 1634 meters above sea level (m.a.s.l), and Maderas, with 1394 m.a.s.l. Figure 2 shows a satellite image of the island of Ometepe, appearing in the upper right part of the image. The shape of the island clearly shows the shape of two peaks formed by volcanoes, Maderas being located on the right and Concepcion on the left. The largest concentration of the population that occupies the island is located in the surroundings of Concepcion and the economy of the island revolves around the tourism, taking advantage of some archeological sites of the pre-Columbian era.

Table 1. Search space for the case study.

| COMPONENT | UNIT OF MEASURE | QUANTITIES IN SEARCH SPACE |
|---|---|---|
| Enercon E-53 WT | Wind Turbine [Rated Power 800 kW] | 0; 1; 3; 4; 5; 6; 7; 8; 9; 12; 15; 18; 21; 24; 27; 30; 33; 36 |
| Solar Park | Installed MW | 0; 2; 4; 6; 8; 10; 12; 14; 16; 18; 21; 24; 27; 30 |
| Converter (representing installed hydropower capacity) | Installed MW | 0; 5.74; 6.90; 8.60; 10.33; 11.48; 13.78; 17.22 |
| Maderas' active volume of PSH reservoir | Batteries [1 battery = 2000 m$^3$] | 0; 1; 3; 6; 9; 12; 15; 18; 21; 27 |
| Geothermal | Installed MW | 0, 1; 2; 3; 4; 5; 6 |

According to Ranaboldo et al. [44], most of the past efforts in Nicaragua related to electrification of areas far from the main cities were focused on grid extensions. However, for many parts of the country, such solutions are technically and financially unfeasible due to the remote and dispersed nature of small communities, or because of geography features that pose a major obstacle to the extension of the electric grid (mountains, forests, lakes, etc.).

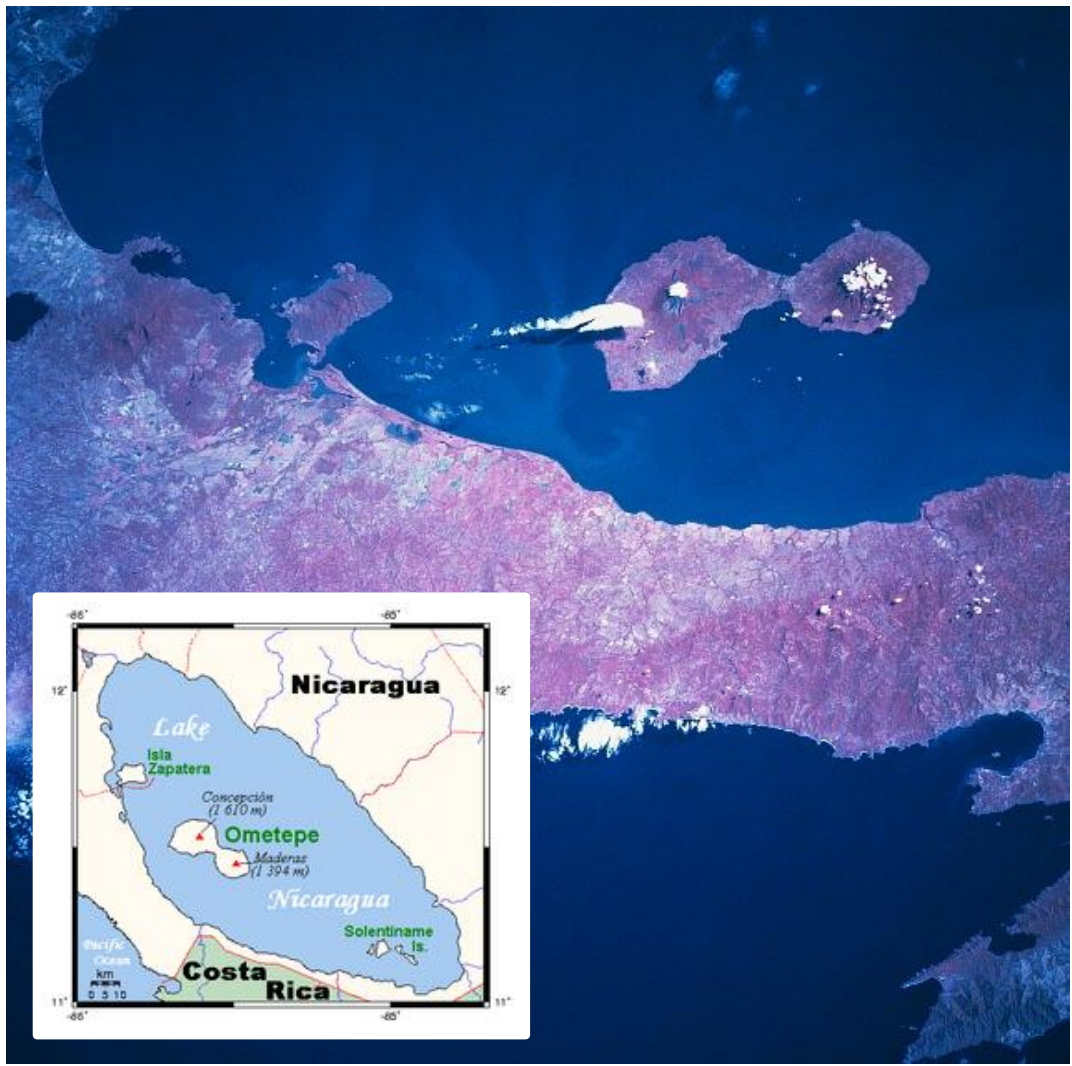

Figure 2. Satellite image showing the region of the island of Ometepe in Nicaragua.

The reminder of this section is composed of six subsections, now dedicated to a better detailing of the components of the system in HOMER, describing the photovoltaic modules, wind turbines, PSH plant, geothermal plant and load profile, finishing with a last subsection mentioning additional considerations.

Based on Figure 1, the hybrid system considers the possible installation of PV modules, wind turbines of the Enercon E-53 type, a geothermal generating plant and a PSH plant to meet a total demand of 70 MWh per day, with a maximum demand value of 4 MW. The PSH plant is modeled in this system by the combination of a battery and a converter, reserving the DC bus for this purpose only.

*3.1. Solar resource*

HOMER requires solar resource inputs for calculating the PV array power output for each hour of the day and year. Based on the latitude and longitude, HOMER is able to retrieve monthly solar data for the specified latitude and longitude from NREL's (National Renewable Energy Laboratory) and NASA's (National Aeronautics and Space Administration) satellite databases. Using this HOMER capacity, Figure 3 presents the results for the daily radiation (in kWh/m²/d) at Ometepe.

By 2014, solar accounted for only 0.10% (1.4 MW) of the total installed capacity of the SIN was solar [42]. Even with great potential for solar energy production, PV arrays in Nicaragua are almost entirely used for self-generation in households and small industry. In more recent years, the most important solar project in Nicaragua is Polaris Solar Park, with 12 MW of installed capacity, which was completed in November 2017 [44].

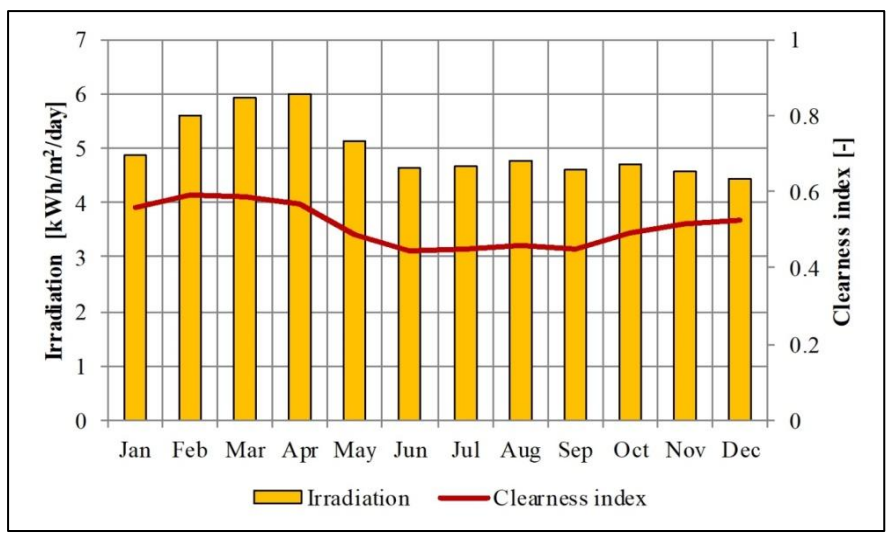

**Figure 3.** Solar resource inputs for Ometepe.

In order to limit the land-use requirements for the solar power plant, the maximum size considered for the solar park was set at 30 MW. According to the information available at [45], the land requirements for a PV solar park could range between 0.89 to 4.90 ha/MW, with a capacity-weighted average of 2.80 m²/MW. Ometepe total area is 276 km² (27,600 hectares) [47], meaning than in the worst-scenario, the land use requirements would be 150 hectares (0.53% of the island area).

*3.2. Wind turbine and wind speed data*

For wind power estimation, HOMER can make use of time series data with 8,760 values representing the average wind speed, in meters per second, for each hour of the year. The wind resource for Ometepe was obtained from the Modern-Era Retrospective Analysis for Research and Applications, Version 2 (MERRA-2), which is a reanalysis dataset produced by Global Modeling and Assimilation Office (GMAO) at NASA [48]. The average wind data used in HOMER for each hour of the year was the average for the corresponding hour from January 1998 to April 2018. Based on these considerations and information, the wind resource input used by HOMER is shown in Figure 4.

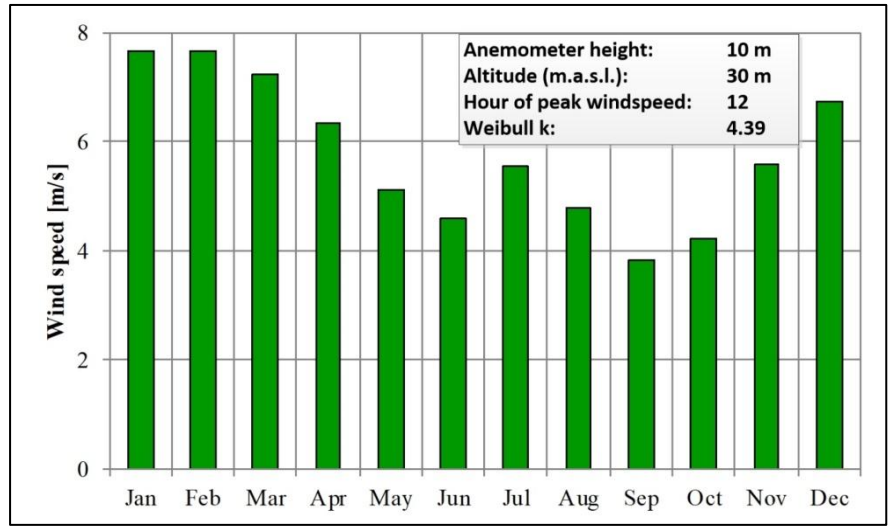

**Figure 4.** Wind resource inputs for Ometepe.

Ometepe is one of the regions with significant wind power potential in Nicaragua [49]. However, the only relevant wind parks of the country are located in Rivas, a few kilometers (~30 km) from the island. Amayo wind farm, which started operating Phase I in 2009, and Phase II in 2010, has an installed capacity of 63 MW, corresponding to 30 SUZLON S88 wind turbines of 2.1 MW each [50]. An initial set of simulations were completed in HOMER and, for the wind regime found by using data from MERRA-2, it was found that better power output from the available wind in Ometepe could be achieved by using ENERCON E-53 wind turbines (800 kW rated power), described in [51], thus, this type of turbine is used when evaluating this case study. The turbine could be placed offshore the island, in the shallow waters of lake Nicaragua.

*3.3. Representing pumped storage hydropower*

Crater lakes situated in dormant or extinct volcanos usually have fresh water. The existing crater lake at Maderas volcano (11°26′44″ N, 85°30′54″ W) will be used as the upper reservoir of the PSH plant. This volcano is a small, asymmetrical stratovolcano that, along with the Concepción volcano and connected by the Istián isthmus, forms the dumbbell-shaped Ometepe island located in Lake Nicaragua, Nicaragua. According to studies and evidence, Maderas volcano has been inactive for tens of thousands of years and future volcanic eruptions are not considered a threat [46]. Lake Nicaragua, with an elevation around 32 m.a.s.l. and average depth of 13 m, will serve as the lower reservoir.

According to [52], Maderas' crater lake volume is approximately 239,000m$^3$, with the water surface ranging within 1203 and 1214 meters above sea level. Based on the information described, it was considered that the available head, taken as a fixed value, is H = 1170 m, both for pumps as for turbines. From [53], some pumped storage plants with a head above 1000 m can be found in Austria, France, Italy, Romania and Spain.

As stated by [34], a combination of a battery bank and a converter can be used in order to represent a PSH plant in Homer. For estimating the PSH cost, where important savings can be achieved by the fact that no dam would need to be built, the set of cost functions presented in [54], which were created from real information for Small Hydropower Projects in Nicaragua, was used for this purpose. As shown in Table 1, this case study considers several options for the installed capacity of the PSH plant. The conversion efficiency, both for pumps and for turbines, is assumed at 85%, an acceptable value according to [15].

For the upper reservoir, the authors of this paper considered that the lake could never be totally emptied, thus the dead storage volume is the one corresponding to the height 1203 m (~4578 m$^3$). Based on that, the unitary battery (equivalent) for this case study is the one corresponding to the volume between heights 1203 m and 1204 m (~2000 m$^3$). Using this volume, H = 1170 m as available head, and 10 kV as reference voltage in Eqs. (2) and (3), it produces an $E_S$ = 6.33 MWh and $C_B$ = 633 Ah, resulting in a stored energy of 3.2 kWh/m$^3$. The maximum charge current is set to 1,377 A, equivalent to a maximum installed pumping capacity of 13.7 MW with 85% efficiency and a maximum pump flow $Q_P$ = 1.2 m$^3$/s. This pump flow would refill from dead storage level to the maximum active level in 12 hours. The storage capacity assessed in this paper ranges within 1 (Z = 1204 m; Vol ≈ 2000 m$^3$) and 27 (Z = 1212 m; Vol ≈ 54,000 m$^3$) equivalent batteries.

Even if there is no need to build a dam, and with a good extent of forestation in the Maderas' volcanic cone, it is important to mention that there are some geological hazards on this volcano, like earthquakes and lahars [46], meaning that sound geological studies, an adequate design of the water intake, and reinforcing the stability of some areas of the mountainside could be required in order to implement this technical solution.

*3.4. Geothermal*

Because of its several lagoons and lakes, and the chain of volcanoes that runs along the Pacific region of the country, Nicaragua is commonly referred to as "the land of lakes and volcanoes". The 16 active volcanoes, crater lakes, volcanic calderas, and numerous areas with hydrothermal activity, are clear indicators of magmatic bodies with high geothermal potential. By 2016, the estimated geothermal potential for Nicaragua was 1197 MW, from which 146 MW were identified for Ometepe island [55].

For estimating the capital cost per installed kW, it was used the average price of US$5,140/kW for geothermal projects in Nicaragua found in [56], which is within the range of cost around the world shown in [15]. The O&M costs are considered proportional to the energy production. Following the remarks described in [57], this O&M cost was set at HOMER as US$20/MWh.

Based on the information found at the at the electricity map site [40] , the average geothermal $CO_2$ emissions used in this paper are set at 38 g$CO_2$/kWh.

*3.5. Load*

The AC load profile used in this case study was created from load curves of years 2014 and 2015, related to hourly demand along the day [58] and monthly demands along the year [42] in Nicaragua. The annual average demand per person in Nicaragua is approximately 580 kWh/y [59]. Based on these load profiles and annual demand, the latter were scaled, and a synthetic data series was created for its HOMER use, based on the corresponding monthly and hourly scaled demand and standard deviation, in order to generate the 8760 hours of energy demand for a population in Ometepe island of 44,000, according to [46]. The hourly demand along the day and monthly demands along the year are shown in Figure 5.

Based on the information found at the electricity map site [40], and using information from July 12th, 2018 to July 15th, 2018 as sample, the average $CO_2$ emissions of the SIN was estimated as 316 g$CO_2$/kWh (standard deviation = 48 g$CO_2$/kWh). The average cost of energy for this period was $0.123/kWh.

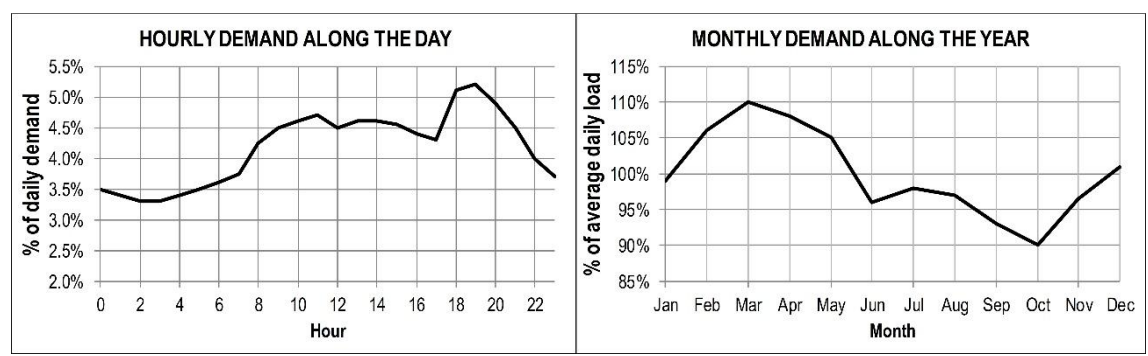

**Figure 5.** Scaled load profiles for the case study.

*3.6. Additional considerations*

Besides the information previously described for the resources and components of the system, there are some additional considerations and assumptions that must be included in the HOMER model in order to run the simulation for the proposed 100% renewable system. The lifetime of the project and equipment was assumed as 20 years, a common assumption for electromechanical equipment, with an annual real interest rate of 6%. As for the Maderas' crater lake, the upper reservoir of the PSH represented by the equivalent battery bank, the technical lifetime for this component was set as 60 years, based on [15]. The lifetime for the steam turbines of the geothermal plant, expressed in operating hours, was set at 100,000 hours, following what is mentioned in [60]. When HOMER calculates the NPC, and according to Lambert et al. [39], if the upper reservoir lifetime is 60 years, the salvage value of this component will be two-thirds of its capital cost at the end of the 20-year project lifetime.

As for the system constraints, the arrangement must count with enough spare capacity to serve sudden increases in the load. HOMER includes this consideration in the operating reserve parameter, and the 10% default value is used. The load-following strategy was adopted as the dispatch policy, and the maximum acceptable capacity shortage value was set at zero, meaning that electricity shortfalls are not allowed to happen.

As stated before, one aim of this paper is to assess the impact of the generation cost in the complementarity between solar and wind power generation and optimal configuration of the system. Because of that, some important assumptions made for modeling this case study in HOMER are:

- The system will be assessed for two conditions for serving base load: using geothermal power and using only VRES (wind and solar).
- The operating reserve for solar and wind power was set at 20% for both. This means that the hybrid power system must keep sufficient spare capacity in operation to attend the load even if the output suddenly increases 20%. In addition to serve as spinning reserve, the importance of this operating reserve is supported by the fact that accurately forecasting renewables is difficult. For example, even though is possible to forecast wind power output a day in advance, forecast errors of 20% - 50% are not uncommon [11].
- To perform the sensitivity analysis related to the impact of cost per installed kW on the system configuration, the capital cost for installed kW, both for PV panels as for wind turbines, was set at US$2000/kW, and the multipliers of this cost used as sensitivity values were 0.75 and 1.25, meaning that the costs range within US$1500/kW and US$2500/kW, acceptable values for renewable projects in Nicaragua [42, 49, 61, 62].
- For estimating the O&M cost, it was assumed that the annual O&M cost equals 3% of the capital cost for wind turbines and solar park, based on [63]-[64]. For PSH, represented in the HOMER model as the combination of the battery bank and the converter, the O&M cost was set at 6% of the capital cost, according to the remarks made by [65]. This O&M cost is independent of the sensitivity values applied to the capital cost for PV and Wind power installed kW.

## 4. Simulation results and discussion

Based on the method and case study previously described, this section presents and discusses the main results of the simulation and sensitivity analysis made with HOMER. Besides the sensitivity analysis related to the impact of cost per kW installed on the hybrid system configuration, this paper evaluated the importance of a reliable and continuous energy source (i.e.: geothermal power) to serve the base load.

For all conditions described for the case study, HOMER was able to find a feasible system configuration. This suggests that is technically feasible to supply Ometepe island with electricity coming entirely from its renewable energy resources.

This section has three subsections. The first presents results related to systems including the geothermal alternative, while the second subsection deals with systems that do not include it. A third subsection includes some additional considerations that are not related only to these two possibilities.

*4.1. System configuration when considering geothermal option*

When the existence of a geothermal plant for serving base load is considered, the optimal system configuration is almost uniform for all sensitivity analysis cases. The results are summarized and shown in Table 2. As expected, due to its dispatchable nature and based on the constraints and operating conditions previously described, geothermal power provides the capacity for serving the baseload for the island. For all sensitivities regarding solar and wind power capital cost, wind was always part of the optimal system configuration found by HOMER, energy storage is required for all cases, and solar power was never in the mix. Intermediate and peak loads are supplied by the intermittent source and PSH , limiting the excess electricity, and minimizing the size and quantity of wind turbines, required hydropower and energy storage. The excess electricity values shown in Table 2 include the electricity required to pump water to refill the upper reservoir.

Figure 6 shows the optimization space for the hybrid systems that consider the geothermal power plant option. Even if they were not specified in the sensitivity analysis, HOMER also calculated results for PV capital cost multipliers lower than those shown in Table 2. Optimal solutions close to multiplicative factors equal to one do not include photovoltaic modules. But there is an inclined dividing line, between approximately the values of 0.35 and 0.60 for the PV capital cost multiplier, from which the solutions begin to include them. These amounts correspond to relatively low acquisition costs that can be reached in eventual situations like the purchase of large quantities of PV panels or with some kind of governmental financial support.

**Table 2.** Summarized results for the sensitivity analysis considering geothermal power plant.

| PV Capital Cost (USD/kW) | | $2000 | $2000 | $2000 | $1500 | $1500 | $2500 | $2500 |
|---|---|---|---|---|---|---|---|---|
| Wind Power Capital Cost (USD/kW) | | $2000 | $1500 | $2500 | $2000 | $2500 | $2000 | $1500 |
| Description | Units | 1 | 2 | 3 | 4 | 5 | 6 | 7 |
| Primary Load | MWh/d | **69.92** | **69.92** | **69.92** | **69.92** | **69.92** | **69.92** | **69.92** |
| PV | MW | - | - | - | - | - | - | - |
| E-53 Units (800 kw each) | units | 4 | **5** | _3_ | 4 | **5** | _3_ | 4 |
| Maderas active volume | m$^3$ | **4000** | **4000** | **4000** | **4000** | **4000** | **4000** | **4000** |
| Geothermal power installed | MW | **3** | **3** | **3** | **3** | **3** | **3** | **3** |
| Converter | MW | **5.74** | **5.74** | **5.74** | **5.74** | **5.74** | **5.74** | **5.74** |
| Total Capital Cost | x10^6 USD | **27.36** | _26.96_ | _26.96_ | **27.36** | _26.96_ | _26.96_ | **27.36** |
| Total NPC | x10^6 USD | 37.86 | _36.03_ | **39.12** | 37.86 | _36.03_ | **39.12** | 37.86 |
| Total O&M Cost | x10^6 USD/year | 0.86 | 0.85 | 0.86 | 0.86 | 0.85 | 0.86 | 0.86 |
| Cost of Energy | USD/kWh | 0.13 | 0.12 | 0.13 | 0.13 | 0.12 | 0.13 | 0.13 |
| Geothermal Production | GWh/yr (% of total) | 16.61 (61.6%) | _14.05_ (52.1%) | **19.16** (71.2%) | 16.61 (61.6%) | _14.05_ (52.1%) | **19.16** (71.2%) | 16.61 (61.6%) |
| Wind Production | GWh/yr (% of total) | 10.33 (38.4%) | 12.92 (47.9%) | 7.75 (28.8%) | 10.33 (38.4%) | 12.92 (47.9%) | 7.75 (28.8%) | 10.33 (38.4%) |
| Tot. Electrical Production | GWh/yr | 26.94 | 26.97 | 26.91 | 26.94 | 26.97 | 26.91 | 26.94 |
| AC Primary Load Served | GWh/yr | 25.51 | 25.51 | 25.51 | 25.51 | 25.51 | 25.51 | 25.51 |
| Excess Electricity | % | 5.3% | 5.4% | 5.2% | 5.3% | 5.4% | 5.2% | 5.3% |
| Battery Throughput | MWh/yr | 4415.05 | 4391.14 | 4336.81 | 4415.05 | 4391.14 | 4336.81 | 4415.05 |
| Battery Autonomy | hours | 4.35 | 4.35 | 4.35 | 4.35 | 4.35 | 4.35 | 4.35 |

Even when it's possible to serve the load entirely with geothermal, the optimal configuration found by HOMER always included wind power and energy storage from PSH. This supports what is mentioned in [12], that in most cases, geothermal plants are more economical serving base load.

13 / 19

Using the sensitivity case scenario where the capital cost favors the installation of 5 wind turbines, Figure 7 shows the average monthly electrical production. The total installed capacity for this case, not including PSH turbines, is 7MW (the peak load for the system is ≈4MW). From December to April, the wind speed regime at Ometepe would allow to serve the most part of the load using Wind Power. For this scenario, the average capacity factor for the wind turbines is 36.9%, and 53.5% for geothermal, with this latter required to work 4700 hours per year in average, allowing to extend the operational life of the steam turbines beyond the project lifetime. The estimated cost of electricity (COE, expressed in $/kWh) for this system configuration, and under the assumed conditions, would be around US$0.13/kWh, including energy storage. This capacity factors and COE are within the range of acceptable values for electricity production from renewables, according to [12] and are even better than the simulated costs for renewables in Nicaragua found in [49]. Nevertheless, it is important to remark that the geothermal potential in Ometepe is on reconnaissance status and has not been completely assessed [55], therefore, more feasibility studies are required in order to provide a more accurate COE for this alternative.

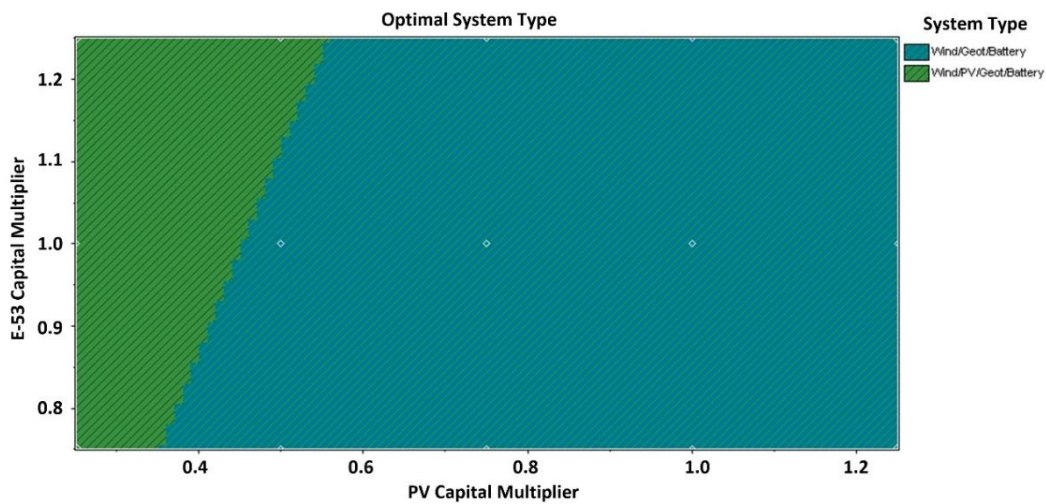

**Figure 6.** Optimization space for the systems with geothermal power plant.

Because of the small size required for the upper reservoir of the PSH, the environmental and economic aspects would be extremely important to determine if PSH is the better option for energy storage and backup power for this system configuration, instead of actual battery banks or using fuel plants for serving intermediate and peak loads. Alternatively, if PSH is found feasible, a greater active volume of the upper reservoir could provide stability in the electricity supply for future population growth or intense economic development of the island.

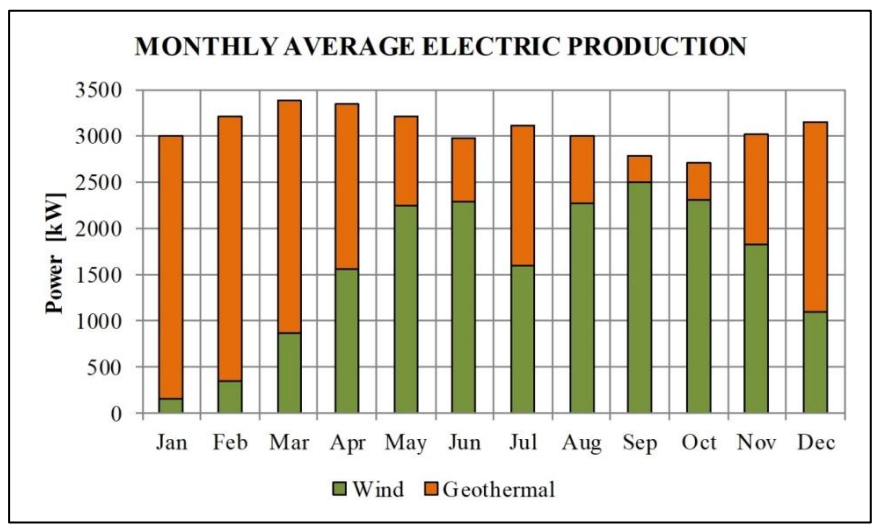

**Figure 7.** Monthly average electric production for system configuration with geothermal plant.

Another beneficial impact of adopting a 100% non-combustion renewable power system for Ometepe island would be the CO2 emission reduction. When considering the inclusion of geothermal power and based on information available at [40], it might be possible to avoid the emission of more than 6,000 tons/year of $CO_2$. Based on the $38 gCO_2/kWh$ considered by the electricity map site, the estimated CO2 emissions for the geothermal plant is less than 700 tons/year. Nevertheless, it is important to mention that the CO2 emissions for geothermal power of this study could range between 180 tons/year and 7200 tons/year if the reference values shown at [12] for tropical areas are considered, meaning that an adequate design and O&M of the technology used is required. This becomes even more relevant if carbon credits are part of the economic analysis.

*4.2. System configuration without geothermal option*

When the geothermal option is not included, the generation costs from VRES have significant impacts on the system components. Based on the described conditions and sensitivities, the most important outcomes of the sensitivity analysis are shown in Table 3. The optimization space in this case indicates that a combination of both VRES and the PSH plant are always part of optimal solution, as shown in Figure 8.

**Table 3.** Summarized results for the sensitivity analysis without geothermal power plant.

| PV Capital Cost (USD/kW) | | $2000 | $2000 | $2000 | $1500 | $1500 | $2500 | $2500 |
|---|---|---|---|---|---|---|---|---|
| Wind Power Capital Cost (USD/kW) | | $2000 | $1500 | $2500 | $2000 | $2500 | $2000 | $1500 |
| Description | Units | 1 | 2 | 3 | 4 | 5 | 6 | 7 |
| Primary Load | MWh/d | 69.92 | 69.92 | 69.92 | 69.92 | 69.92 | 69.92 | 69.92 |
| PV | MW | 10.00 | 8.00 | 14.00 | 14.00 | 14.00 | 10.00 | 8.00 |
| E-53 Units (800 kw each) | units | 24 | 27 | 15 | 15 | 15 | 24 | 27 |
| Maderas active volume | $m^3$ | 18000 | 18000 | 30000 | 30000 | 30000 | 18000 | 18000 |
| Converter | MW | 5.74 | 5.74 | 6.90 | 6.90 | 6.90 | 5.74 | 5.74 |
| Total Capital Cost | x10^6 USD | 70.94 | 60.94 | 77.19 | 64.19 | 70.19 | 75.94 | 64.94 |
| Total NPC | x10^6 USD | 99.67 | 89.94 | 105.17 | 92.17 | 98.17 | 102.80 | 93.94 |
| Total O&M Cost | x10^6 USD/year | 2.50 | 2.53 | 2.44 | 2.44 | 2.44 | 2.50 | 2.53 |
| Cost of Energy | USD/kWh | 0.34 | 0.31 | 0.36 | 0.32 | 0.34 | 0.36 | 0.31 |
| PV Production | GWh/yr (% of total) | 14.83 (19.3%) | 11.87 (14.5%) | 20.77 (35.0%) | 20.77 (35.0%) | 20.77 (35.0%) | 14.83 (19.3%) | 11.87 (14.5%) |
| Wind Production | GWh/yr (% of total) | 62 (80.7%) | 69.75 (85.5%) | 38.75 (65.0%) | 38.75 (65.0%) | 38.75 (65.0%) | 62 (80.7%) | 69.75 (85.5%) |
| Tot. Electrical Production | GWh/yr | 76.84 | 81.62 | 59.52 | 59.52 | 59.52 | 76.84 | 81.62 |
| AC Primary Load Served | GWh/yr | 25.51 | 25.51 | 25.51 | 25.51 | 25.51 | 25.51 | 25.51 |
| Excess Electricity | % | 66.8% | 68.7% | 57.1% | 57.1% | 57.1% | 66.8% | 68.7% |
| Battery Throughput | MWh/yr | 19.56 | 19.56 | 32.59 | 32.59 | 32.59 | 19.56 | 19.56 |
| Battery Autonomy | hours | 1245.06 | 982.12 | 2581.42 | 2581.42 | 2581.42 | 1245.06 | 982.12 |

The high excess energy results for all system configurations (>50%) suggests that there is little complementarity in time between solar and wind resources, meaning that their availability often occur simultaneously. For example, it is worth noticing that the hour of peak wind speed is at noon (See Figure 4). This little complementarity increases the required size of the solar and wind parks, thus increasing the cost of electricity. The minimum installed capacity, excluding the PSH, is 26MW, 6.5 times the peak load, for a system configuration including a 14 MW solar park and fifteen E-53 wind turbines. For this system configuration, the capacity factors estimated by HOMER were 16.9% for the PV panels and 36.9% for the wind park.

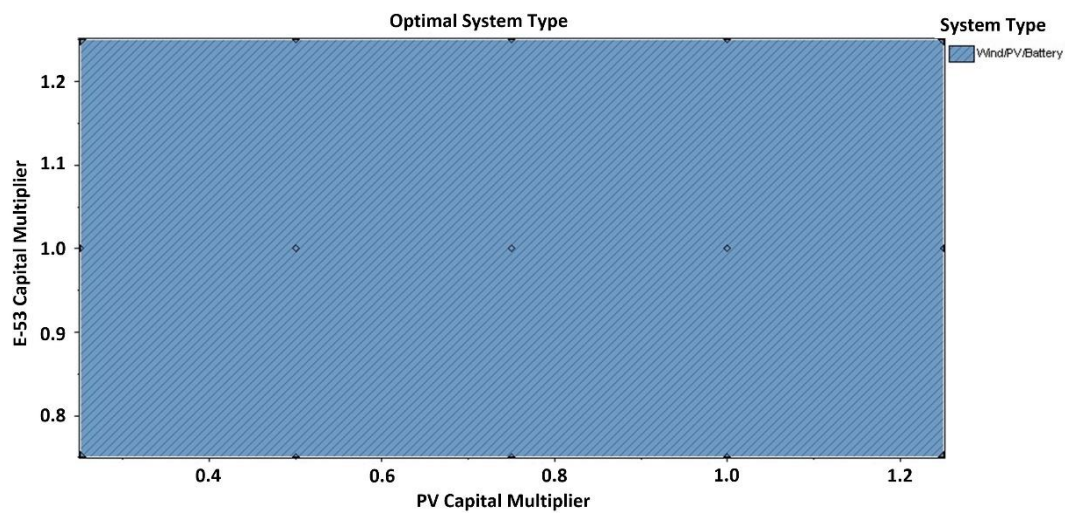

**Figure 8.** Optimization space for the systems without geothermal power plant.

Based on the results shown in Table 3, and for the conditions evaluated and described in the case study, the most economic configuration, in terms of NPC and COE was system #2, and the most expensive was system #3. In terms of COE, the values ranged within US$0.31/kWh and US$0.36/kWh. According to Asturias [49], the generation cost in Nicaragua for non-combustion renewables varies between $0.08/kWh (hydropower minimum cost) and $0.28/kWh (wind power maximum cost). However, it is important to remember, that for this work, the lifetime of the project and the PV panels and wind turbines was set as 20 years, and no sensitivity analysis was conducted on the O&M cost. These two parameters have an important impact on the COE of each system configuration [38].

*4.3. Additional remarks*

Besides being required in all system configurations and for all sensitivities, energy storage through the PSH system using the Maderas' crater lake and lake Nicaragua as reservoirs poses an interesting option in terms of energy storage economy. For this case study, the balance provided by higher energy storage capacities corresponded to smaller total electricity production. It is important to point out that all the optimal solutions provided by HOMER and discussed in the two previous subsections include the PSH plant, regardless of the inclusion of a geothermal power plant or photovoltaic panels.

To offset the high cost of a system completely based on VRES, the excess electricity could be sold to the grid (Ometepe is 30 km far from the city of Rivas), or could be used in activities that do not require on-demand power and are not affected by intermittency, for example, water purification, wastewater treatment, air conditioning or irrigation pumps [66-67]. Drinking water supply problems in Ometepe are not uncommon, as evidenced in [68-69]. Another interesting option to consider, both for social and for economic development of the island, is to improve water quality of lake Nicaragua and commercial fish production through aeration using this excess energy, which could translate into positive environmental, social and economic impacts that may well justify these investments through government subsidies.

Finally, the most important comment that can be made based on the results, is that the sensitivity analysis performed using the HOMER software capacities proved that the economic aspects of electricity production from renewable energy resources are at least as important as the energy resources availability itself. Obviously, the availability of renewable resources is necessary to meet a demand with this type of resource, but the technical and economic feasibility of the use of these resources is also necessary.

## 5. Conclusions

The aim of this paper was to evaluate how the capital cost of renewable technologies affects the optimal configuration and cost of electricity of an isolated electric power system, comprising only non-combustion RES (geothermal, wind, solar, PSH). Ometepe island, in Nicaragua, was used as case study, especially because it is located in a sweet water lake, has geothermal potential and the crater lake at the top of one of its volcanoes makes it an interesting option for considering PSH. The methods described and used in this paper are based on Canales and Beluco [34] and the recommendations provided by the user support at HOMER Energy website [35]. The simulation, optimization and sensitivity analysis were performed in the Legacy Version of HOMER Energy software [37].

The obtained results are in line with the general comments found in the literature. When geothermal power is considered, this technology is able to serve the base load of the system, reducing the installed capacity of other renewables, as well as decreasing the storage requirements and excess electricity production. This configuration also evidenced that, for the case study, wind power is a more reliable and effective energy source than solar power. Shifting to a 100% non-combustion renewable power system drastically reduces the carbon footprint of the power generation of the system.

When the geothermal option is not included, and the intermittent renewable resources have low complementarity in time, the required size of the solar and wind parks is enormous, consequently rising the COE and excess electricity production.

For the conditions evaluated in the case study the optimum solution for the hybrid system including the geothermal plant would include four wind turbines, an upper PSH reservoir with a capacity of 4,000 m3 and a geothermal power plant with a capacity of 3 MW. The capital cost for this system would be US$ 37,723,856 with COE of US$ 0.129 per kWh. The optimum solution for the hybrid system without the geothermal plant would inclued 24 wind turbines, an upper PSH reservoir with capacity of 18,000 m3 and a 10 MW solar park, for a capital cost of US$ 98,168,616 and COE of US$ 0.366 per kWh.

The different hybrid system configuration results demonstrated that economic aspects of renewable energy generation (e.g.: different capital costs combinations) are at least as important as the natural resources availability. Also, regarding economic aspects of this work, Ometepe was selected as case study because the use of a crater lake as the upper reservoir for PSH significantly reduces infrastructure investment. All the optimal system configurations found by HOMER required energy storage capacities.

This paper assessed Ometepe island as an isolated system, in order to evaluate its capacity to become completely independent from the SIN. Future studies could well evaluate how the RES available at Ometepe could integrate within this system and how other storage devices can be used in hybrid storage systems with the Maderas PSH plant.

Besides a deeper evaluation (including technical and environmental aspects) of the PSH plant feasibility using the Maderas' crater lake as upper reservoir, a possible future research is how excess energy from VRES can be sold to the SIN or used to improve the overall life quality of the island.

### Acknowledgments

The authors are grateful for the support received by their institutions for the research work that resulted in this paper. The third author acknowledges the financial support received from CNPq for his research work (proc. n.312941/2017-0).

### Conflicts of Interest

The authors declare no conflict of interest.